# A systematic construction of Gaussian basis sets for the description of laser field ionization and high-harmonic generation

Aleksander P. Woźniak,[1,a)] Michał Lesiuk,[1] Michał Przybytek,[1] Dmitry K. Efimov,[2,3]
Jakub S. Prauzner-Bechcicki,[4] Michał Mandrysz,[2] Marcelo Ciappina,[5,6,7] Emilio Pisanty,[5,8]
Jakub Zakrzewski,[2,9] Maciej Lewenstein,[5,10] and Robert Moszyński[1]

AFFILIATIONS

[1] Faculty of Chemistry, University of Warsaw, Pasteura 1, 02-093 Warsaw, Poland
[2] Institute of Theoretical Physics, Jagiellonian University in Krakow, Łojasiewicza 11, 30-348 Kraków, Poland
[3] Department of Theoretical Physics, Faculty of Fundamental Problems of Technology, Wrocław University of Science and Technology, 50-370 Wrocław, Poland
[4] Marian Smoluchowski Institute of Physics, Jagiellonian University in Krakow, Łojasiewicza 11, 30-348 Kraków, Poland
[5] ICFO—Institut de Ciencies Fotoniques, The Barcelona Institute of Science and Technology, Av. Carl Friedrich Gauss 3, 08860, Castelldefels, Barcelona, Spain
[6] Physics Program, Guangdong Technion-Israel Institute of Technology, Shantou 515063, China
[7] Technion-Israel Institute of Technology, Haifa 32000, Israel
[8] Max Born Institute for Nonlinear Optics and Short Pulse Spectroscopy, Max-Born-Straße 2A, Berlin 12489, Germany
[9] Mark Kac Complex Systems Research Center, Jagiellonian University in Krakow, Łojasiewicza 11, 30-348 Kraków, Poland
[10] ICREA, Pg. Lluís Companys 23, 08010 Barcelona, Spain

a) Author to whom correspondence should be addressed: awozniak@chem.uw.edu.pl

**ABSTRACT**

A precise understanding of mechanisms governing the dynamics of electrons in atoms and molecules subjected to intense laser fields has a key importance for the description of attosecond processes such as the high-harmonic generation and ionization. From the theoretical point of view, this is still a challenging task, as new approaches to solve the time-dependent Schrödinger equation with both good accuracy and efficiency are still emerging. Until recently, the purely numerical methods of real-time propagation of the wavefunction using finite grids have been frequently and successfully used to capture the electron dynamics in small one- or two-electron systems. However, as the main focus of attoscience shifts toward many-electron systems, such techniques are no longer effective and need to be replaced by more approximate but computationally efficient ones. In this paper, we explore the increasingly popular method of expanding the wavefunction of the examined system into a linear combination of atomic orbitals and present a novel systematic scheme for constructing an optimal Gaussian basis set suitable for the description of excited and continuum atomic or molecular states. We analyze the performance of the proposed basis sets by carrying out a series of time-dependent configuration interaction calculations for the hydrogen atom in fields of intensity varying from $5 \times 10^{13}$ W/cm$^2$ to $5 \times 10^{14}$ W/cm$^2$. We also compare the results with the data obtained using Gaussian basis sets proposed previously by other authors.

Published under license by AIP Publishing. https://doi.org/10.1063/5.0040879

## I. INTRODUCTION

Attoscience is a rapidly developing area of research with opportunities of unprecedented applications in physics, chemistry, and biology. In attoscience, the basic process consists of shining a short (a few optical cycles long) pulse of near- or mid-infrared laser radiation on a target, which can be an atom, a molecule, a biomolecule, etc.[1–3] The response of the target may result in high-harmonic generation (HHG)[4–11] and, if the intensity of the pulse is high enough, in electron detachment and formation of atomic or molecular ions.[12–30] The high harmonics are generated as a part of a macroscopic phase-matched process[31] and manifest themselves as extreme ultraviolet (XUV) or x-ray pulses at frequencies given by integer multiples of the driving pulse frequency.[32–36] These pulses, although not very intense, have attosecond-scale duration and exhibit a very high spatial and temporal coherence,[37–40] so they allow for imaging of the structure and dynamics of matter in their natural length and time scales.

The HHG spectrum is characterized by its distinctive shape. For a few lowest harmonic peaks, the intensity decreases with the harmonic order according to lowest-order perturbation theory.[41] Afterward, a long plateau is observed in which the harmonic peaks are of a similar height. This plateau terminates with a sharp cutoff, beyond which the emission decays exponentially. This shape is usually explained using a semiclassical picture, encoded in the famous three-step model of HHG.[7–9] In the first step, the electron escapes the potential well of the nucleus via tunneling ionization. In the second step, it is further accelerated in the laser field until the direction of the field changes. Finally, in the third step, it is reaccelerated toward the residual ion and recombines with it, which is accompanied by an emission of a high-energy photon. The maximum energy of the emitted photons, and thus the position of the harmonic cutoff, depends on the ponderomotive energy $U_p$ and on the ionization potential $I_p$ via the relation $E_{\text{cutoff}} = I_p + 3.17 U_p$. For inversion-symmetric media, such as atoms in the gas phase, only peaks at odd harmonics are present in the spectrum due to symmetry constraints.

Theoretical description of the HHG spectra requires the knowledge of the time evolution of the dipole moment in the non-perturbative regime. This is a formidable task, since, in turn, it requires solving the time-dependent Schrödinger equation (TDSE). The latter problem is usually treated in two ways: either by diverse analytical or semi-analytical methods, such as the probably most famous strong field approximation (SFA),[42] or by using a variety of purely numerical methods.[43] Among the former, one of the most predominant is the previously mentioned three-step model by Lewenstein et al.[7–9,42] It is able to describe a broad range of one- and two-electron processes, including the HHG, the above-threshold, tunneling and non-sequential ionization, and the electron rescattering, but the results it provides are mostly qualitative. In order to obtain more quantitative insight, several more accurate alternatives have been proposed in the literature, such as the R-matrix theory[44] and the Floquet state analysis.[45–48]

Regarding the numerical techniques for solving the TDSE, recent developments in the propagation of the wavefunction in real-time on a finite spatial grid have proven extremely useful for a very precise description and understanding of the electron dynamics during intense field processes.[49–63] However, this is applicable mostly to one-electron systems, such as the hydrogen-like ions and the $H_2^+$ ion, since full-dimensional numerical integration of TDSE containing an explicit two-electron wavefunction remains challenging.[64–71] Extending this approach to more complex systems is barely feasible, as it requires huge amounts of memory and computational time and must rely on using various approximations, such as the most commonly used single-active-electron approximation[6,72–83] or restricted dimensionality models,[84–95] recently extended to the three-electron case.[96,97] Such techniques may be useful for some general insight into the electron dynamics but fail to capture more subtle effect occurring in the HHG, as pointed out by Gordon et al.[98] Due to these limitations, novel methods that can describe many-electron systems are still in high demand.

A kind of a "third way" to describe the attosecond processes may be to apply the methods widely used in quantum chemistry that employ the expansion of the wavefunction of the examined system in a basis set of predefined functions, most commonly the Gaussian-type functions or Gaussian-type orbitals (GTOs). Currently, for the majority of the widely used quantum chemistry methods, their real-time time-dependent counterparts have been developed and applied to the real-time propagation, including time-dependent multi-configurational Hartree–Fock,[99–104] time-dependent configuration interaction (TDCI),[105–111] time-dependent density functional theory,[104,112–115] algebraic diagrammatic construction,[116,117] and time-dependent coupled cluster.[118–120] This so-called basis set approach is much less limited by the number of electrons and atoms than the grid-based one. It is also computationally less expensive, since most of the necessary calculations can be performed analytically. However, the biggest shortcoming of this approach is the deficiency of basis sets to describe the motion of electrons in high-energy fields. Since most of the quantum chemical studies are focused on determining the properties of atoms and molecules in their ground states, a great majority of the existing atomic basis sets, such as Pople, Dunning, or Ahlrichs basis sets, have been obtained by minimizing the atomic ground state energies at a suitable level of theory (usually, the Hartree–Fock or a simple post-HF method).[121] It is thus natural that these basis sets approximate the lowest-energy states much more accurately compared to the higher excited states. When subjected to an intense laser field, the atoms or molecules usually become excited to states near or above the ionization threshold. What we need, then, are basis sets that are able to describe a sufficiently large number of excited and continuum atomic or molecular states, with a precision comparable to that of the description of the ground state.

It is quite obvious that the construction of a Gaussian basis set that correctly mimics the Rydberg continuum orbitals is far from trivial, mostly due to their highly oscillatory behavior. One of the very first attempts was made by Kaufman et al. in 1989.[122] They presented a method for constructing a set of Gaussian-type orbitals (GTOs),

$$\chi^{\text{GTO}}_{lm;\alpha}(r, \theta, \phi) = \sqrt{\frac{2(2\alpha)^{l+3/2}}{\Gamma(l + \frac{3}{2})}} r^l \exp(-\alpha r^2) Y_{lm}(\theta, \phi), \quad (1)$$

with the exponents $\alpha$ fitted to maximize their overlaps with a series of the so-called Slater-type orbitals (STOs),

$$\chi^{\text{STO}}_{nlm;\zeta}(r, \theta, \phi) = \frac{(2\zeta)^{n+1/2}}{\sqrt{\Gamma(2n+1)}} r^{n-1} \exp(-\zeta r) Y_{lm}(\theta, \phi), \quad (2)$$

with a constant exponent $\zeta$ and a variable positive integer principal quantum number $n$ [$Y_{lm}(\theta, \phi)$ are the real spherical harmonics]. In principle, the idea behind this is highly appealing. STOs constitute a complete basis set in the Hilbert space (as opposed to the hydrogenic orbitals), making possible the description of the electronic continuum without the need to use any non-square-integrable functions. They also possess some properties particularly useful from the perspective of real-time propagation: their radial functions are identical for all angular momenta $l$ and they have an expectation value of $r$ given by

$$\langle \chi^{STO}_{nlm;\zeta} | r | \chi^{STO}_{nlm;\zeta} \rangle = \frac{2n+1}{2\zeta}. \quad (3)$$

Since in the real-time-dependent methods, the wavefunction is most often propagated in a finite region of space (usually accompanied by some kind of absorbing boundary conditions), one can estimate *a priori* how many STO shells are needed in order to describe the wavefunction up to the boundaries of this region. On the other hand, due to their exponential character, STOs lack some features unique to the Gaussian-type functions, such as an efficient evaluation of multicenter integrals. In their paper, Kaufmann *et al.* have shown that their functions (referred to as the Kaufmann functions or K functions further in the text) are able to generate a discretized spectrum of the continuum eigenfunctions and imitate the Coulomb wave functions up to considerably long distances away from the nuclei.[122]

Unfortunately, the K functions also possess some major drawbacks. First, since each STO is described with a single GTO, the size of the basis set scales linearly with the number of STOs to be reproduced. Moreover, the high-$n$ STOs are approximated less accurately compared to the low-$n$ ones. Finally, the STOs form a non-orthogonal basis set, and the overlap integral between two adjacent STOs approaches unity with the increase in $n$. It therefore follows that for large $n$, the values of the exponents of the K functions also become closer to each other, creating a risk of linear dependencies appearing in the basis set, which may jeopardize the numerical stability of the calculations.

The disadvantages listed above make the Kaufmann basis sets rather ineffective for the description of states close to or beyond the ionization threshold, the description of which may require including STOs with very high principal quantum numbers. Therefore, their applicability in attoscience is limited to simulating atoms and molecules in laser fields of relatively low intensities. Due to the same reason, the K functions may be inefficient for the description of heavier elements. Since the STO's principal quantum number can be associated with the atomic shell number, for atoms with a large number of occupied shells, it may turn out that the linear dependencies and fitting inaccuracies prevent one from representing more than a few lowest excited states accurately.

It is worth noting that after the paper by Kaufmann *et al.*, several attempts to create Gaussian basis sets suitable for the description of the excited and continuum states have appeared. Nestmann and Peyerimhoff reported fitting linear combinations of GTOs to spherical Bessel functions for the purpose of the electron–molecule scattering calculations.[123] Their work was later extended to the Coulomb wave functions by Faure *et al.*[124] Moiseyev and co-workers used Gaussian-type and Hylleraas-type functions to describe the high harmonic generation in the helium atom in long laser pulses using the Floquet approach.[46,48,125] Some more elaborate approaches, such as the B-spline basis sets or combinations of Gaussians and grid-based methods, have also been employed with moderate success in describing both ionization rates and HHG.[126–131] Fiori and Miraglia[132] and later Szczygieł *et al.*[133] explored the approximation of the continuum wave functions of the hydrogen atom with plane wave functions multiplied by the Gaussian-type orbitals (GTOPWs or London orbitals), which can mimic the oscillatory behavior of continuum orbitals. Such basis sets were able to very accurately reproduce the measured ionization spectra of the hydrogen and helium atoms as well as molecular-frame photoelectron angular distributions for the hydrogen molecule[132,133] but failed to exceed beyond the perturbative regime. Rowan *et al.* showed that only their counterparts with time-dependent parameters are able to reasonably describe the wavefunction propagation in laser fields.[134]

Recently, Coccia and co-workers have decided to revisit and extend the work of Kaufmann *et al.* with the primary goal of constructing basis sets for an accurate description of atomic and molecular HHG spectra.[131,135–139] They combined the K functions with augmented Dunning basis sets[140] containing a very large number of diffuse functions. Their method proved useful for the description of HHG at intensities below the barrier-suppression ionization threshold. However, the limitations stemming from the use of the K functions, such as near-linear dependencies within the basis set, could not be avoided.

In this paper, we introduce a novel systematic scheme for constructing finite Gaussian basis sets for an optimal representation of both bound and continuum Hamiltonian eigenstates, designed particularly to enable efficient time-dependent calculations of many-electron and multicenter systems. Recalling the benefits of the Kaufmann approach, such as simplicity and physical interpretability of the K functions, we determine a series of Gaussian functions to best reproduce a given subset of STOs. However, our approach is somewhat different. Instead of optimizing a single GTO for each consecutive STO, we optimize all GTOs at the same time so that every STO from a predefined range may be approximated by their linear combination with a roughly equal accuracy.

This paper is organized as follows: In Sec. II, we briefly discuss the theory behind the basis set approach to real-time propagation. The procedure used to obtain optimal basis sets is described in Sec. III, while the computational details of time-dependent calculations are presented in Sec. IV. Although the basis sets designed by us are to be used eventually for simulations of many-electron systems, we optimize them by performing time-dependent configuration interaction (TDCI) calculations of the hydrogen atom in intense laser fields, as it is an example of a system for which a reliable and accurate numerical (grid-based) reference is easily obtainable. The results of the simulations are presented and discussed in Sec. V, as well as compared with the results obtained using basis sets composed of the K functions. We also present some preliminary results for the helium atom as a simplest example of a many-electron system. Finally, in Sec. VI, we conclude our work.

## II. OUTLINE OF THE THEORY

As stated in Sec. I, the main goal of the real-time propagation is to solve the time-dependent Schrödinger equation,

$$i\frac{\partial}{\partial t}|\Psi(\mathbf{r},t)\rangle = \hat{H}(t)|\Psi(\mathbf{r},t)\rangle, \quad (4)$$

where both the Hamiltonian $\hat{H}(t)$ and the electronic wavefunction $|\Psi(\mathbf{r},t)\rangle$ depend explicitly on time. In most numerical approaches, it is assumed that the total evolution time can be discretized, i.e., divided into a large number of small yet finite time steps $\Delta t$ during which the time-dependency of the Hamiltonian can be ignored, and the formula for a finite difference propagator is introduced,

$$|\Psi(\mathbf{r},t+\Delta t)\rangle = U(t+\Delta t,t)|\Psi(\mathbf{r},t)\rangle, \quad (5a)$$

$$U(t+\Delta t,t) = \exp(-i\Delta t H(t+\Delta t/2)). \quad (5b)$$

The fractional time step was introduced by using the midpoint rule when approximating the integral of the Hamiltonian from $t$ to $t + \Delta t$, which improves the stability of the algorithm and reduces the error. The length of a single time step must be chosen carefully, as the proper choice strongly depends on the rate of change of the Hamiltonian and therefore on the frequency of the laser field $\omega$. In our calculations, we have assured that $\Delta t \ll 2\pi/\omega$ and checked the convergence with respect to $\Delta t$.

In most cases, the Hamiltonian can be expressed as a sum of a time-independent part $H_0$, describing the unperturbed atom or molecule, and a time-dependent light–matter interaction operator $V(t)$, $\hat{H}(t) = \hat{H}_0 + \hat{V}(t)$. In the TDCI formalism, we expand the time-dependent wavefunction as the combination of ground and excited states of the system, obtained through diagonalization of the time-independent Hamiltonian matrix in the basis set of choice,

$$|\Psi(\mathbf{r},t)\rangle = \sum_{n\geq 1} c_n(t)|\psi_n(\mathbf{r})\rangle. \quad (6)$$

From now on, the position dependence of states will be omitted as obvious. We are thus able to represent the wavefunction as a column matrix $\mathbf{c}(t)$ of the time-dependent coefficients $c_n(t)$ and reformulate the propagation equation (5) as a matrix equation

$$\mathbf{c}(t+\Delta t) = \exp(-i\Delta t[\mathbf{H}_0 + \mathbf{V}(t+\Delta t/2)])\mathbf{c}(t) \quad (7)$$

with matrix elements $(H_0)_{ij} = \langle\psi_i|\hat{H}_0|\psi_j\rangle$ and $(V(t))_{ij} = \langle\psi_i|\hat{V}(t)|\psi_j\rangle$. It is also worth noting that in the case of one electron systems, such as the hydrogen atom, the states $|\psi_n\rangle$ reduce to linear combinations of basis functions, $\chi_k|\psi_n\rangle = \sum_{k\geq 1} c_k\chi_k$.

The two key quantities obtained from the real-time propagation are the HHG spectrum and the ionization probability. In this work, the former is calculated in the dipole form as the squared modulus of the Fourier transform of the dipole moment expectation value,

$$I_{\text{HHG}}(\omega) = \left|\frac{1}{t_f - t_i}\int_{t_i}^{t_f}\langle\Psi(\mathbf{r},t)|\hat{d}|\Psi(\mathbf{r},t)\rangle e^{i\omega t}dt\right|^2, \quad (8a)$$

$$\langle\Psi(\mathbf{r},t)|\hat{d}|\Psi(\mathbf{r},t)\rangle = \sum_{i\geq 1}\sum_{j\geq 1}[c_i(t)]^\dagger c_j(t)\langle\psi_i|\hat{d}|\psi_j\rangle, \quad (8b)$$

where we integrate over the total propagation time. We prefer to use the dipole form instead of the, also frequently used, acceleration form. This is because the dipole acceleration can be computed analytically at each time step only using the Ehrenfest theorem, which holds true only for exact wavefunctions. As this paper focuses on comparing basis sets, which provide different levels of approximation to the exact hydrogen wavefunction, this might be source errors in the comparison. In our calculations, we implement a complex absorbing potential (CAP), the construction and properties of which will be elaborated in Sec. IV. Its presence causes the norm of the excited wavefunction to decrease over time, simulating the ionization process, and thus allows for determining the ionization probability $W(t)$,

$$W(t) = 1 - \langle\Psi(\mathbf{r},t)|\Psi(\mathbf{r},t)\rangle. \quad (9)$$

## III. BASIS SETS

Given the advantages of the Slater-type orbitals set out in Sec. I, we attempt to find a set of Gaussian-type orbitals able to reproduce a given number of STO shells. However, in order to avoid the disadvantages of the Kaufmann method, our scheme will not rely on approximating every STO separately. Instead, we will generate a sequence of GTOs such that each STO from a predefined range can be approximated by their linear combination with a roughly equal precision. Concurrently, we also aim at keeping the basis set free from the linear dependencies.

Similar to the Kaufmann approach, we choose the overlap integral as a criterion of similarity between an STO and a GTO, since it can be associated with the $L^2(\mathbb{R}^3)$ metric in the Hilbert space. By using the orthonormality property of the spherical harmonics, the overlap between two normalized functions of the same $l$—a GTO with exponent $\alpha$ and a STO with exponent $\zeta$ and a principal quantum number $n$—can be evaluated analytically from the following equation:

$$S(n,l,\zeta,\alpha) = \frac{2^{-l/2-1/4}\alpha^{-n/2-l/2-1}\zeta^{n+1/2}}{\sqrt{(2n)!}}$$
$$\times \sqrt{\frac{\alpha^{l+3/2}}{\Gamma(l+3/2)}}\Gamma(n+l+2)\, U\left(\frac{n+l+2}{2},\frac{1}{2},\frac{\zeta^2}{4\alpha}\right), \quad (10)$$

where U($a$, $b$, $z$) is the confluent hypergeometric function of the second kind.[141] The STO exponent may be viewed as an effective nuclear charge seen by the electron. Since in this paper, we focus on the hydrogen atom, from now on, we set $\zeta = 1$.

We start our procedure by defining the reference subset {**S**} of Slater-type orbitals, which we aim at approximating by the Gaussian functions. The choice of this subset depends both on the system under consideration and the simulation conditions. Since the basis set is meant to describe primarily the excited and continuum states, a reasonable choice for the principal quantum number $n$ of the first STO in the subset {**S**} is the number corresponding to the lowest unoccupied atomic shell. The upper bound of the reference subset is less obvious to define, as it should correspond to the highest energy states achievable by the electron in the applied electric field. An exemplary way to estimate it is to take the STO shell with

⟨r⟩ equal to or slightly exceeding the electron's quiver amplitude, defined as $E_0/\omega^2$, where $E_0$ is the maximum amplitude of the electric field.

Since the Gaussian exponents are real positive numbers, a brute-force optimization of the whole basis set seems rather impossible. However, we can discretize the range of possible exponents and construct the so-called sampling set, from which we will select functions to be included in the final basis set. This discretization relies on generating a large set of even-tempered Gaussian (ETG) functions {**G**} (Gaussian functions with constant ratios between adjacent exponents). The ETG functions possess a useful property of spanning the Hilbert space evenly[142] and are therefore able to approximate any function with accuracy dependent solely on the ratio between exponents. The optimal size and range of the sampling set will be discussed later.

The selection of the exponents from the sampling set {**G**} is performed separately for each angular momentum. For each $l$ ranging from $l = 0$ to $l = L_{max}$, we calculate a $S \times G$ matrix of overlap integrals between STOs from the reference subset {**S**} and GTOs from the sampling set {**G**}. First, we need to reject GTOs that have no significant contribution to the description of any STO. We accomplish this by removing the GTOs for which the maximum component of the overlap vector is smaller than the so-called overlap cutoff. Next, we calculate the sums of the components of the remaining vectors (their $L^1$ norms), obtaining for each GTO a quantity that we will refer to as the cumulative performance score,

$$\text{CPS}_i = \sum_{j=1}^{S} |\langle \chi_j^{\text{STO}} | \chi_i^{\text{GTO}} \rangle|. \quad (11)$$

The cumulative performance score may be seen as a measure of an overall performance of a single GTO in the description of all the STOs. The choice of $L^1$ norm is based mostly on the intuitiveness of the results because in this way, the cumulative performance score can take on values from 0 (for functions that poorly approximate the STOs or approximate only a few of them) to $S$ (for functions that are most essential for approximating a majority of STOs). Alternatively, using the $L^2$ norm or the $L^0$ norm (measured as the number of components larger than a predefined threshold) may also be considered.

Next, we start an iterative procedure of selecting GTOs based on their cumulative performance score.

1. From the remaining GTOs, we choose the one with the highest cumulative performance score and include it in the final basis set.
2. Because the initial sampling set should be rather extensive in order to approximate the continuous spectrum of exponents as closely as possible, we usually encounter a large number of GTOs, which overlap with the reference subset in a similar manner to the GTO chosen in step 1. In other words, their overlap vectors are close due to similar values of the exponents. The sampling set must be further depleted of these functions before selecting the next GTO in order to keep the final basis set free of linear dependencies. We achieve this by determining the so-called cosine distance between the overlap vectors of the GTOs. The cosine distance between vectors **A** and **B** is defined as

$$D_{\cos}(\mathbf{A}, \mathbf{B}) = 1 - \cos(\mathbf{A}, \mathbf{B})$$
$$= 1 - \frac{\mathbf{A} \cdot \mathbf{B}}{\|\mathbf{A}\|_2 \|\mathbf{B}\|_2}. \quad (12)$$

Its value may range from 0 to 2; however, in our case, the maximum value is 1, since the overlap integrals are non-negative. When the cosine distance is 1, the overlap vectors are orthogonal, meaning that the function described by the vector **A** overlaps with different STOs than the function described by the vector **B**. When the cosine distance equals 0, the overlap vectors are parallel (differing only by a proportionality constant), meaning that both functions approximate the same STOs, so one of the functions may be eliminated as redundant. In our procedure, we first calculate the cosine distances between the overlap vector corresponding to the function chosen in step 1 and the overlap vectors of all the remaining GTOs.

3. Next, we introduce the so-called cosine cutoff, a value that determines the maximum acceptable similarity between two overlap vectors (thereby also between two basis functions). The GTOs for which the cosine distance calculated in step 2 is smaller than the cosine cutoff are removed.

Steps 1–3 are repeated until all of the GTOs are either removed or included to the final basis set.

Since the core element of our method is defining the range of states that are energetically accessible by the electron, for a basis set constructed according to the above scheme, we propose the name of active range-optimized (ARO) basis set.

The most critical factor in constructing an optimal ARO basis set is the proper choice of the sampling set, i.e., the range of the exponents and the sampling density (the ratio between adjacent exponents). If this range is too narrow or ill-placed, some of the STOs, especially with the lowest or highest $n$, may not be described properly due to the lack of sufficiently diffuse or tight functions. Choosing a range that is too broad should not affect the final outcome, since any redundant functions will be removed anyway due to the overlap cutoff condition, but it may unnecessarily extend the computational time. Since in our scheme, each STO is approximated by a linear combination of GTOs, some functions with relatively large and small exponents may actually be beneficial for a more accurate description. Therefore, a reasonable guess for the sampling set range should be the range covering the exponents of the Kaufmann functions fitting the same reference subset (for the hydrogen atom, approximately from $10^{-4}$ to $10^0$), but extended on both sides by one or two orders of magnitude. The appropriateness of this choice can easily be verified by checking if at least a few of the functions are discarded due to the overlap cutoff condition. As we mentioned in the step 2 of the iterative procedure, the sampling density should be high enough to provide convergence with respect to the number of functions in the final ARO basis set and the values of the selected exponents. In our calculations, about 10 000 ETG functions with exponents ranging between $10^{-k}$ and $10^{-k+1}$ usually proved sufficient to achieve this convergence.

The overlap cutoff governs the number of "enhancing" functions mentioned above, i.e., the tight and diffuse GTOs that usually do not have any major contribution to the description of STOs but merely improve the GTO approximation. The proper choice of the overlap cutoff provides the optimal effective range of exponents from which the functions are selected. Our calculations have shown that the range of values between 0.1 and 0.2 is appropriate.

The cosine cutoff determines the overall number of functions in the final basis set. As previously stated, it also serves as a tool for reducing the linear dependencies in the basis set. It is important to emphasize that these features are achieved not by limiting the similarity of the Gaussian functions themselves, but by limiting the similarity with which two GTOs describe the reference subset. This unique property is the source of the characteristic structure of the ARO basis sets. Any variation in the exponent of a Gaussian function leads to changes in the overlap integrals and, therefore, to a change in the orientation in the overlap vector by a certain angle. In our procedure, the cosine of this angle is compared to the cosine cutoff in order to decide whether a certain function from the sampling set is to be removed or kept for further selection. However, it is interesting to note that the value of this angle depends not only on the variation in the exponent but also on the values of the overlap vector components. Functions that have a large contribution to the majority of the reference subset usually correspond to overlap vectors with most of the components significantly larger than zero. Therefore, a change in an exponent, even a relatively small one, results in variation of the majority of the overlap vector components. The combination of small variations in a large number of components may already suffice to fulfill the cosine cutoff condition. On the contrary, the "enhancing" functions usually correspond to overlap vectors with most of the components close to zero, and thus, a small variation in the exponent changes only a few components of the overlap vector. It is therefore clear that a much larger variation needs to be applied in order to alter the orientation of the overlap vector by an angle sufficient to fulfill the cosine cutoff condition. A direct consequence of this fact is that in the ARO basis sets, the ratios between adjacent exponents are the largest for the tightest and diffuse functions. This ensures the densest distribution of functions that play a key role in approximating the reference subset, as compared to the "enhancing" functions. This feature distinguishes the ARO basis set from the Kaufmann basis set, where the ratios between adjacent exponents decrease with their values, as well as from the ETG basis sets, where this ratio is constant for all adjacent functions.

## IV. COMPUTATIONAL DETAILS
### A. Simulation conditions

We performed a series of TDCI propagations of the hydrogen atom wavefunction in short (femtosecond-scale) pulses of an intense linearly polarized laser field represented by an oscillating electric field. The motion of the electron is considered in the Born–Oppenheimer approximation, while the interaction between it and the external field is treated in the dipole approximation. All calculations are performed in the velocity gauge, as it requires lower angular momenta included in the basis set for the results to converge than in the length gauge.[143] The interaction operator $V_{ext}(t)$, coupling the electron and the electric field polarized along the $z$-axis, reads $V_{ext}(t) = -iA(t)\frac{\partial}{\partial z}$ in the velocity gauge, where $A(t)$ is the vector potential related to the field $E(t)$ by the relation $E(t) = -\partial_t A(t)$. In our calculations, the vector potential corresponds to a time-dependent electric field representing a laser pulse with a sine-squared envelope,

$$E(t) = \begin{cases} E_0 \sin(\omega_0 t) \sin^2(\omega_0 t/2n_c) & \text{if } 0 \leq t \leq 2\pi n_c/\omega_0, \\ 0 & \text{otherwise,} \end{cases} \quad (13)$$

where $E_0$ is the field amplitude related to the laser peak intensity $I_0$ by $I_0 = \epsilon_0 c E_0^2$, $\omega_0$ is the carrier frequency, $n_c$ is the number of optical cycles the pulse consists of, and $2\pi n_c/\omega_0$ is the total duration of the pulse.

All calculations are performed for $\omega_0 = 1.55$ eV ($\lambda = 800$ nm), corresponding to a Ti:sapphire laser, frequently used in the attosecond experiments.[1] The number of optical cycles is either 4 or 20 (corresponding to time intervals of ∼441.3 a.u. and 2206.6 a.u.). The barrier-suppression ionization threshold of the hydrogen atom (the intensity sufficient for the electron to classically overstep the potential barrier generated by the nucleus) is about $1.37 \times 10^{14}$ W/cm$^2$. We examine four laser intensities: two below the ionization threshold, $5 \times 10^{13}$ W/cm$^2$ and $1 \times 10^{14}$ W/cm$^2$, and two above it, $2 \times 10^{14}$ W/cm$^2$ and $5 \times 10^{14}$ W/cm$^2$.

In order to capture the ionization process, a complex absorbing potential (CAP) of the form $-iV_{CAP}(r)$ is also included in the time-dependent Hamiltonian. The structure of the applied CAP is discussed below.

To sum up, the total expression for the time-dependent Hamiltonian, in atomic units, reads

$$H(\mathbf{r}, t) = -\frac{1}{2}\nabla^2 - \frac{1}{r} - iA(t)\frac{\partial}{\partial z} - iV_{CAP}(r), \quad (14)$$

where $r$ denotes the distance between the electron and the nucleus.

### B. Basis set calculations

The basis set calculations are performed using three different ARO basis sets, constructed according to the scheme presented in Sec. III and fitted to reference subsets of STOs with principal quantum numbers ranging from 2 to either 30, 60, or 90. Each basis set includes functions with angular momenta from 0 (s-type orbitals) to 8 (l-type orbitals). The sampling set is the same for each basis set and each angular momentum and comprises of 10 000 ETG functions, with exponents ranging from $10^{-6}$ to $10^1$. The overlap cutoff was empirically set to 0.15, a value that provided the best results in terms of both the HHG spectra and ionization probabilities (when compared to the numerical reference described below). The cosine cutoff was adjusted separately for each angular momentum in each basis set so that in every basis set, the number of functions with a given $l$ is equal to 19 for $l = 0$ or $20 - l$ for $l \geq 1$, reproducing the correct degeneracy of the unoccupied orbitals of the hydrogen atom. Since the ARO construction scheme aims at describing a large number of excited and continuum states, it is natural that these basis sets lack functions decaying sufficiently fast with the distance from the nucleus that are required to describe the 1s hydrogen ground state (see the supplementary material). To compensate for this, each basis set was supplemented with the 1s orbital from the STO-6G hydrogen basis set.

For comparison, we constructed a Kaufmann basis set, also containing functions with nine lowest angular momenta, fitted to STOs with principal quantum numbers from 2 to 20. The STO-6G hydrogen orbital was also added to this basis set. It can be seen that the number of functions is equal for both kinds of basis sets, which ensures a fair comparison of the results. We extended the results of Kaufmann (who provides the Gaussian exponents up to $l = 5$ only[122]) by finding numerical roots of the derivatives of $S(n, l, \zeta, \alpha)$ [Eq. (10)] with respect to $\alpha$ for different values of $n$ and $l$.

After the diagonalization of the time-independent CAP-free Hamiltonian matrix in each of the constructed basis sets, the real-time propagation is performed, starting from the generated ground state. In the case of the hydrogen atom, due to the cylindrical symmetry of the Hamiltonian, only orbitals with the projection of the orbital angular momentum parallel to the linearly polarized electric field are coupled to it and contribute to the dipole moment. Therefore, in order to speed up the calculations, we use only functions with $m = 0$.

Formally, the propagation scheme presented in Eq. (5b) involves a time-ordered matrix exponential, which can be cumbersome to calculate. To avoid this and to keep the propagation algorithm consistent with the one used in the grid-based calculations, we replace the propagator in Eq. (7) with the Crank–Nicolson propagator,[144]

$$U(t+\Delta t, t) = \left(1 + \frac{i\Delta t}{2}H(t + \Delta t/2)\right)^{-1}\left(1 - \frac{i\Delta t}{2}H(t + \Delta t/2)\right), \quad (15)$$

so the matrix propagation equation changes to

$$\left(\mathbf{I} + \frac{i\Delta t}{2}\mathbf{H}(t + \Delta t/2)\right)\mathbf{c}(t + \Delta t) = \left(\mathbf{I} - \frac{i\Delta t}{2}\mathbf{H}(t + \Delta t/2)\right)\mathbf{c}(t), \quad (16)$$

where $\mathbf{I}$ is the identity matrix. The time step $\Delta t$ is set to 0.01 a.u. and the total propagation time is equal to twice the duration of the pulse.

### C. Grid-based calculations

As a numerical reference, we also performed grid-based real-time propagation of the hydrogen atom using the QPROP software.[145,146] These calculations are also performed within the velocity gauge. In QPROP, the hydrogen atom wavefunction is expanded in partial waves (spherical harmonics multiplied by the corresponding time-dependent radial functions),

$$\Psi(\mathbf{r}, t) = \frac{1}{r}\sum_{l=0}^{L_{max}}\sum_{m=-l}^{m=-l} R_{lm}(r, t) Y_{lm}(\theta, \phi). \quad (17)$$

In our calculations, we expand the wavefunction in $L_{max} = 40$ partial waves, which are sufficient to achieve the convergence of the HHG spectra and the ionization probabilities for all investigated laser intensities (meaning that adding functions with higher angular momenta had no effect on the obtained results). The radial grid extends to 120 bohrs, with the spacing set to 0.1 bohr. The propagation algorithm and the time step and the total propagation time are the same as in the basis set calculations.

### D. The complex absorbing potential

Because the used Gaussian basis sets are not complete, due to a finite number of both angular and radial functions, they are not able to properly describe the wavefunction at an arbitrary distance from the nucleus. This causes unphysical reflections of parts of the wavefunction associated with the continuum eigenstates and their interference with the bound states. A similar problem is encountered in the grid-based approach, where the wavefunction is reflected after reaching the grid boundary. The most common way to avoid such artifacts is to use a complex absorbing potential, which effectively eliminates the components of the wavefunction that travel further away from the center of the system than a predefined distance $r_{CAP}$, simulating the ionization process.[147,148] Since the introduction of the CAP breaks the conservation of the norm of the wavefunction, the time propagation is no longer unitary. One of the most prominent and frequently used CAPs is the potential introduced by Manolopoulos, distinguished by its sound mathematical basis.[149–153] Unfortunately, this potential cannot be applied in any basis set calculations, since it contains a singularity that causes the matrix elements of $V_{CAP}$ to diverge. This difficulty can be overcome by using a polynomial form for the CAP. Thus, in both the basis set calculations and the grid-based calculations, we implement a quadratic CAP of the form

$$V_{CAP}(r) = \eta\,\theta(r - r_{CAP})(r - r_{CAP})^2, \quad (18)$$

where $\theta(r)$ is the Heaviside step function and $r_{CAP}$ denotes the starting position of the CAP. The parameter $\eta = 2.4 \times 10^{-4}$ was optimized to reproduce the results obtained on a grid with the CAP derived by Manolopoulos.[149] The details of the construction of the CAP and its properties are presented in the Appendix.

The choice of the CAP starting position is not unique and it can strongly affect the final results, especially the ionization probability. In some studies, $r_{CAP}$ is defined by the quiver amplitude of the electron in the oscillating electric field and thus depends on the simulation conditions.[77,154,155] In others, it is placed more arbitrarily based on the van der Waals radius of the examined atom.[156] In the present paper, we propose a universal method for determining an optimal $r_{CAP}$, which depends only on the simulated system and not on the external perturbation. It can simultaneously be derived based on three different premises:

1. The first reasoning is purely theoretical. For any atom, we can define a critical electric field strength above which the electron can escape the Coulomb potential of the nucleus in a classical manner, and the ionization mechanism switches from the tunneling ionization to the barrier-suppression ionization.[157] For hydrogen-like ions, the value of this field is usually defined as $E_{crit} = Z^3/16$, so for the hydrogen atom, we obtain $E_{crit} = 0.0625$ a.u., which corresponds to the previously mentioned value of intensity $1.37 \times 10^{14}$ W/cm$^2$. Since the three-step model of HHG involves the tunneling step, we can assume that only the electron trajectories that do not exceed the quiver amplitude associated with $E_{crit}$ contribute to the harmonic generation. We can thus define $r_{CAP}$ as the critical field quiver amplitude, which for the hydrogen atom is equal to 19.3 bohrs.

2. The CAP starting position should be located in a range where the asymptotic ionization probability measured at the end of the time propagation is invariant with respect to $r_{CAP}$. We performed a series of grid-based propagations for the hydrogen atom using our CAP with $r_{CAP}$ varying from 5 bohrs to 35 bohrs and found this range to be about 9 bohrs–21 bohrs. A similar reasoning concerning the invariance of the ionization rates with respect to $r_{CAP}$ was adopted by Sissay *et al.*, who achieved results consistent with ours.[158]

3. The optimal value of $r_{CAP}$ should also maximize the relative height difference between the harmonic plateau and the

background beyond the harmonic cutoff, leading to the sharpest cutoff. Placing the CAP starting position too close from the nucleus eliminates some of the electron trajectories that should end in the recombination and harmonic generation, lowering the intensities of harmonic peaks. Similarly, placing it too far from the nucleus spuriously includes some trajectories of the ionized electron in the HHG process, elevating the post-cutoff background. By performing similar test calculations as in the previous point, we determined the value of $r_{CAP}$ that maximizes the cutoff height to be about 19 bohrs.

It can be seen that all three approaches lead to a similar value of the optimal $r_{CAP}$, which is about 19 bohrs. In our calculations, we extend this value to 19.5 bohrs in order to ensure that the CAP minimally overlaps with the bound eigenstates.

## V. RESULTS

### A. Time-independent calculations

Before comparing the results obtained via the real-time propagation, we first focus on the general properties of the ARO basis sets, particularly their ability to approximate the time-independent Hamiltonian eigenspectrum of the hydrogen atom. This provides us a preliminary assessment of their potential to describe the evolution of the time-dependent wavefunction. The characteristics of each of the constructed ARO basis sets (named ARO30, ARO60, and ARO90 based on the maximum principal quantum number of the Slater orbital in the respective reference subset), compared with the Kaufmann basis set (analogously named K20), are presented in Table I. As predicted, the ARO construction scheme encompasses a much wider range of GTO exponents. This range also shifts toward smaller exponents as the reference subset is extended since the STOs with higher principal numbers are characterized by larger values of $\langle r \rangle$. Due to this fact, the ARO basis sets are practically free of linear dependencies. Of the three examined basis sets, only ARO30 required a removal of five linear combinations of GTOs in order to avoid numerical instabilities during the diagonalization of the Hamiltonian matrix, while in the case of the K20 basis set, this number exceeds 40.

**TABLE I**. The characteristics of the ARO basis sets and the Kaufmann basis set used throughout the calculations.

|  | ARO30 | ARO60 | ARO90 | K20 |
|---|---|---|---|---|
| Number of basis set functions[a] | 144 | 144 | 144 | 144 |
| Number of linearly independent functions[a,b] | 139 | 144 | 144 | 101 |
| Lowest exponent[c] | $6.048\,084 \times 10^{-4}$ | $7.329\,910 \times 10^{-5}$ | $9.702\,192 \times 10^{-6}$ | $1.711\,786 \times 10^{-3}$ |
| Highest exponent[c] | 1.969 985 | 1.229 846 | $8.045\,241 \times 10^{-1}$ | $1.012\,151 \times 10^{-1}$ |
| Number of bound Hamiltonian eigenstates ($E < 0$) | 31 | 47 | 64 | 23 |
| Number of continuum Hamiltonian eigenstates ($E > 0$) | 108 | 97 | 80 | 78 |
| Lowest Hamiltonian eigenvalue (ground state energy) | −0.499 908 | −0.499 907 | −0.499 905 | −0.498 694 |
| Highest Hamiltonian eigenvalue | 21.881 96 | 12.609 12 | 7.715 54 | 0.985 72 |

[a]Only functions with $m = 0$ counted.
[b]With the threshold for the minimal overlap matrix eigenvalue equal to $10^{-8}$.
[c]Not counting the STO-6G 1s orbital.

**TABLE II**. Few lowest orbital energies calculated with the used basis sets. The ellipses denote that none of the computed Hamiltonian eigenvalues can be attributed to a given hydrogenic state.

| Orbital | Exact value | ARO30 | ARO60 | ARO90 | K20 |
|---|---|---|---|---|---|
| 1s | −0.5 | −0.499 908 | −0.499 907 | −0.499 905 | −0.498 694 |
| 2s | −0.125 | −0.124 995 | −0.124 993 | −0.124 987 | −0.124 450 |
| 2p | −0.125 | −0.124 988 | −0.124 988 | −0.124 986 | −0.124 314 |
| 3s | −0.0(5) | −0.055 556 | −0.055 556 | −0.055 556 | −0.055 551 |
| 3p | −0.0(5) | −0.055 554 | −0.055 553 | −0.055 552 | −0.055 359 |
| 3d | −0.0(5) | −0.055 552 | −0.055 552 | −0.055 551 | −0.055 304 |
| 4s | −0.031 25 | −0.031 250 | −0.031 250 | −0.031 250 | −0.031 243 |
| 4p | −0.031 25 | −0.031 250 | −0.031 250 | −0.031 250 | −0.031 240 |
| 4d | −0.031 25 | −0.031 249 | −0.031 249 | −0.031 248 | −0.031 163 |
| 4f | −0.031 25 | −0.031 249 | −0.031 249 | −0.031 248 | −0.031 137 |
| 5s | −0.02 | −0.019 999 | −0.020 000 | −0.020 000 | −0.019 904 |
| 6s | −0.013(8) | −0.013 881 | −0.013 889 | −0.013 889 | −0.012 559 |
| 7s | −0.010 204 | −0.009 933 | −0.010 204 | −0.010 204 | ... |
| 8s | −0.007 813 | ... | −0.007 812 | −0.007 812 | ... |
| 9s | −0.006 173 | ... | −0.006 070 | −0.006 173 | ... |
| 10s | −0.005 | ... | ... | −0.004 992 | ... |

Interestingly, extending the reference subset of the ARO basis sets increases the number of bound Hamiltonian eigenstates at the expense of the number of continuum eigenstates. However, at the same time, the highest Hamiltonian eigenvalue also tends to rapidly decrease, resulting in a denser distribution of the states just above the ionization potential (which should be crucial in simulations with fields close to the ionization threshold). The highest number of continuum eigenstates per unit of energy is still achieved using the K20 basis set, but the number of bound states generated by this basis set is far smaller compared to each of the ARO basis sets.

The energies of the first few hydrogen orbital energies are presented in Table II. In general, each of the ARO basis sets provides more accurate energies of the hydrogen orbitals than the K20 basis set. The degeneracies of the atomic shells are also reproduced more faithfully. The number of correctly described shells increases with the ARO reference subset size, and for the ARO90 basis set, we obtain an excellent agreement with the exact energies up to $n = 10$. It is also worth noting that apart from the increase in accuracy of the high energy states, the accuracy of the lowest eigenvalues tends to decrease slightly. Although seemingly worrying, this may actually indicate that the ultimate purpose and motivation behind the ARO construction scheme—the ability of the basis set to reproduce as many eigenstates as possible with a comparable accuracy—strives to be fulfilled.

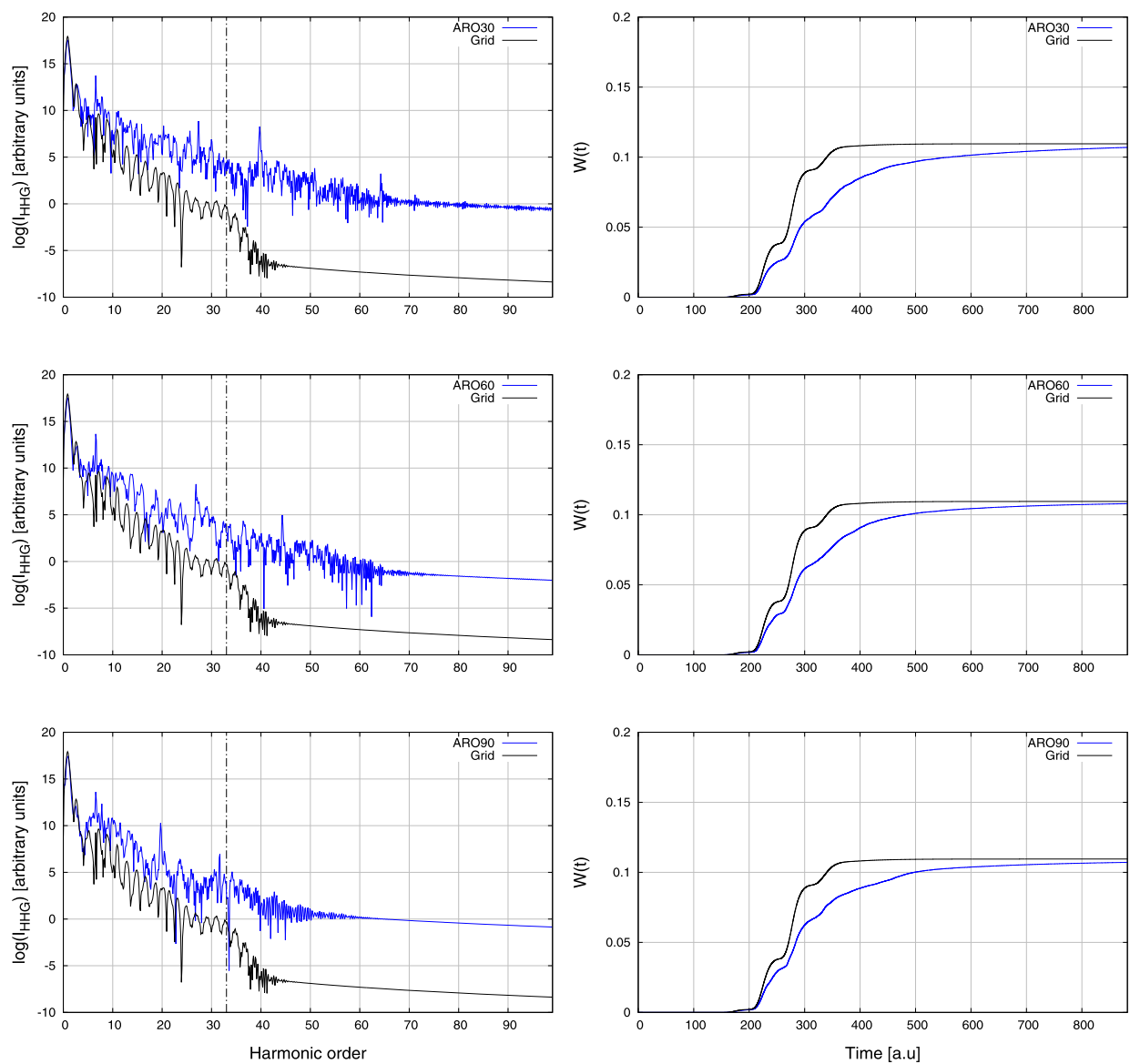

**FIG. 1**. HHG spectra (left column) and ionization probabilities (right column) of the hydrogen atom at $I_0 = 2 \times 10^{14}$ W/cm$^2$ and $n_c = 4$, obtained using three different ARO basis sets, compared with the grid-based results. The HHG spectra are presented up to thrice the harmonic cutoff value (denoted by the dotted-dashed line).

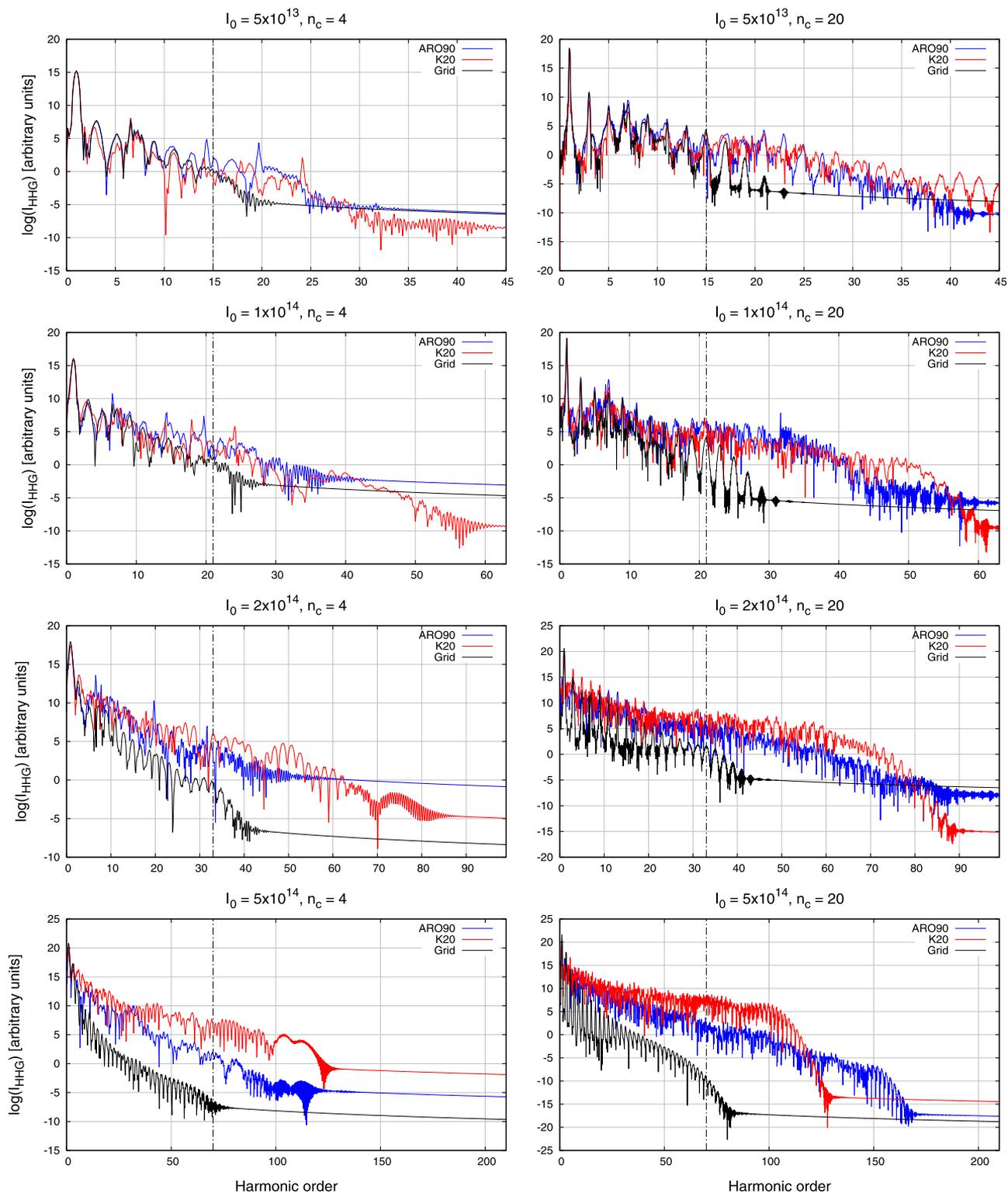

**FIG. 2**. HHG spectra of the hydrogen atom at different laser intensities and different numbers of optical cycles, obtained using the ARO90 basis set, using the K20 basis set, and from the grid-based calculations. The spectra are presented up to thrice the harmonic cutoff value (denoted by the dotted-dashed line).

### B. Real-time propagations

In this subsection, we discuss the quality of the results obtained via the real-time propagation of the hydrogen atom wavefunction using the constructed ARO basis sets. Let us start by analyzing the effect of the size of the ARO reference subset on real-time observables. The exemplary HHG spectra and ionization probabilities calculated using the ARO30, ARO60, and ARO90 basis sets for the case of $I_0 = 2 \times 10^{14}$ W/cm$^2$ and $n_c = 4$ are presented in Fig. 1. It can easily be seen that extending the ARO reference subset crucially improves the quality of the HHG spectrum, reducing the noises in the HHG background beyond the harmonic cutoff. It also has a positive effect on the presence and position of the cutoff itself. In the spectrum obtained using the ARO30 basis set, the cutoff is barely distinguishable, as the intensities of the subsequent harmonic peaks decrease incrementally with the increase in the harmonic order. In the spectrum corresponding to the ARO60 basis set, the cutoff is already visible but located too far compared to the numerical result (between 50th and 60th harmonics). Finally, in the ARO90 spectrum, the cutoff is not only clearly visible but also matches both the grid results and the theoretical value of the 33rd harmonic.

The effect of the size of the reference subset on the ionization probability is less striking, but still visible. It can be seen that along with adding more STOs, the curves become more ridged and the inflection points corresponding to the consecutive optical cycles become more distinct, resembling the grid results.

Since of all the examined ARO basis sets, the ARO90 one provides clearly the best results, for the sake of clarity and transparency in the further discussion, we no longer analyze the results obtained using the ARO30 and ARO60 basis sets.

A comparison of the HHG spectra obtained using the ARO90 basis set, the K20 basis set, and the grid-based calculations for all simulation conditions considered here is presented in Fig. 2. It can be seen that in all cases except $I_0 = 5 \times 10^{14}$ W/cm$^2$ and $n_c = 20$, the ARO90 basis set predicts the cutoff position closer to the theoretical reference than the K20 basis set. The K20 basis set performs particularly underwhelmingly at the intensities below the ionization threshold, as the harmonic peaks are present even after twice the theoretical cutoff position. The ARO90 is also observed to reproduce the shapes of the lowest harmonic peaks more accurately. In general, both basis sets perform worse in the 20 optical cycle cases than in the four optical cycle ones. This results from the fact that during longer pulses, the amount of energy absorbed by the system is larger and the atom becomes excited to higher energy states that may be described less accurately using a finite set of functions. Such artifacts are the shortcoming of the basis set approach in general when compared to the grid methods. However, while the picture provided by the ARO basis sets still has potential to be further improved by extending the reference subset and increasing the number of functions, the K basis sets clearly reach the limit of their possibilities due to growing linear dependencies (that is why we do not compare basis sets containing more functions).

Not to rely solely on the visual assessment of the spectra, we should introduce a quantitative measure that will allow us to quantify how accurately each of the basis sets reproduces the spectrum obtained using the grid-based calculations. No such measure for comparing the HHG spectra has been proposed in the literature thus far, but our suggestion is to treat the HHG spectrum as any other signal and apply tools known from the signal processing. A common method for determining the similarity of two signals relies on the so-called correlation distance, which for signals $\mathbf{A}$ and $\mathbf{B}$ is defined as

$$D_{\text{corr}}(\mathbf{A}, \mathbf{B}) = 1 - \frac{(\mathbf{A} - \bar{\mathbf{A}}) \cdot (\mathbf{B} - \bar{\mathbf{B}})}{\|(\mathbf{A} - \bar{\mathbf{A}})\|_2 \|(\mathbf{B} - \bar{\mathbf{B}})\|_2}, \quad (19)$$

where $\bar{\mathbf{A}}$ and $\bar{\mathbf{B}}$ are the mean values of the signals $\mathbf{A}$ and $\mathbf{B}$. One can immediately notice the resemblance to the previously introduced cosine distance (12). The only difference is that the correlation distance includes the mean values of the signals, and thus, it is insensitive to shifting the signals along the $y$ axis. In comparing the HHG spectra, this is more appropriate, since we are interested only in the relative differences between values at given points on the $x$ axis (e.g., the presence of the harmonic peaks beyond the theoretical cutoff position). Similar to the cosine distance, the correlation distance can also assume values from 0 to 2. For $0 < D_{\text{corr}} < 1$, the signals are positively correlated, for $D_{\text{corr}} = 1$, there is no (either positive or negative) correlation between the signals, and for $1 < D_{\text{corr}} < 2$, the signals are negatively correlated. In our case, $\mathbf{A}$ and $\mathbf{B}$ are the decimal logarithms of the HHG spectra and $\bar{\mathbf{A}}$ and $\bar{\mathbf{B}}$ are their mean values. The reason for such a choice is that the intensity of the lowest harmonics is usually larger by several orders of magnitude than the intensity of the highest ones, so it assures that the correlation distance depends on the whole shape of the spectrum and not only on the shapes of the few highest peaks. The correlation distances between the spectra obtained from the basis set calculations and from the grid-based calculations are presented in Table III.

**TABLE III.** The correlation distances between the HHG spectra obtained from the basis set calculations and from the grid-based calculations under different simulation conditions. When calculating the correlation distance, the spectra were truncated at the harmonic order equal to thrice the harmonic cutoff.

| $I_0$ | $n_c$ | ARO90 | K20 |
| --- | --- | --- | --- |
| $5 \times 10^{13}$ | 4 | 0.067 75 | 0.098 79 |
|  | 20 | 0.160 25 | 0.221 43 |
| $1 \times 10^{14}$ | 4 | 0.061 79 | 0.132 08 |
|  | 20 | 0.199 53 | 0.219 76 |
| $2 \times 10^{14}$ | 4 | 0.044 91 | 0.118 20 |
|  | 20 | 0.152 15 | 0.312 49 |
| $5 \times 10^{14}$ | 4 | 0.075 07 | 0.137 54 |
|  | 20 | 0.167 56 | 0.220 46 |

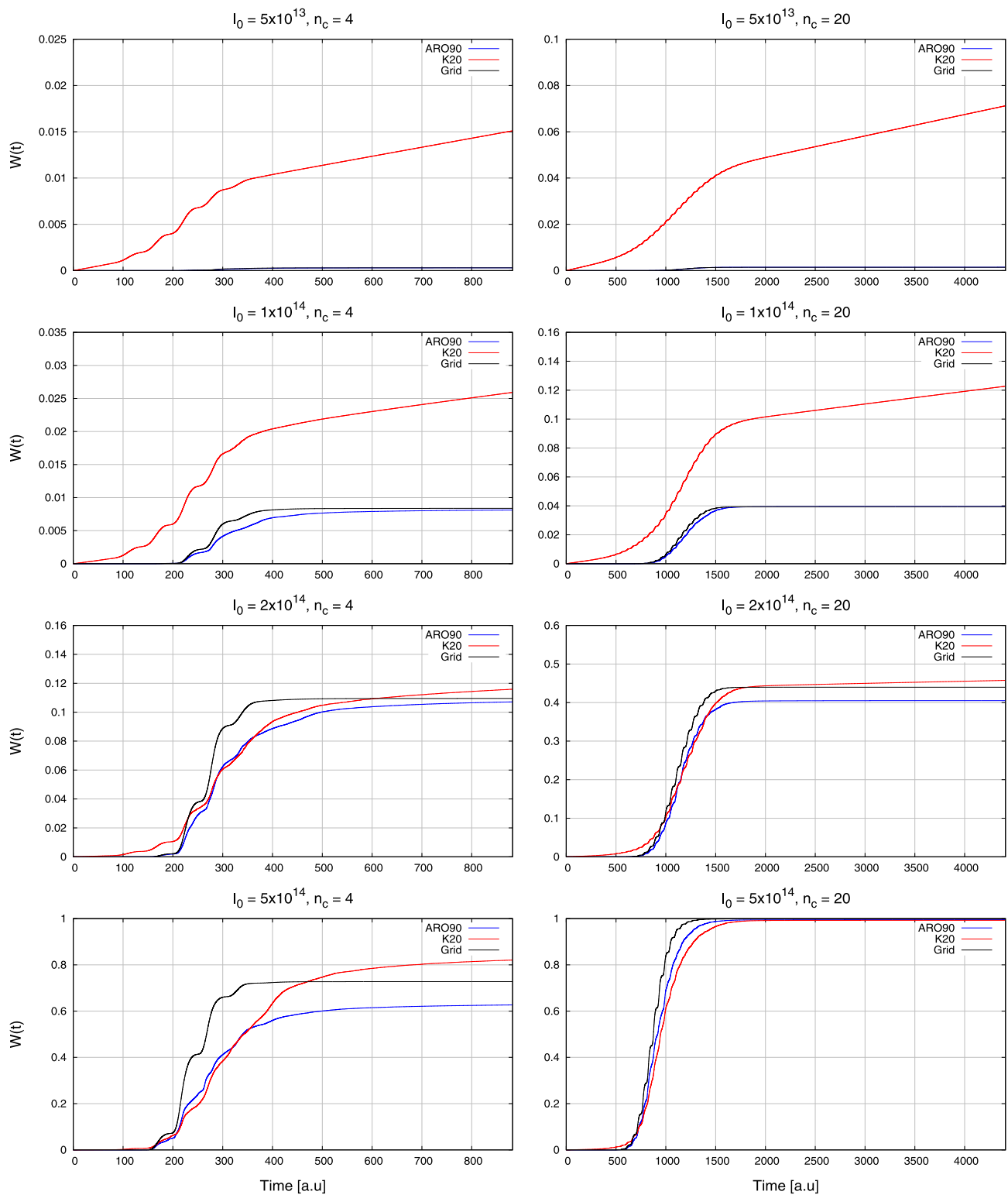

**FIG. 3**. Ionization probabilities of the hydrogen atom at different laser intensities and different numbers of optical cycles, obtained using the ARO90 basis set, using the K20 basis set, and from the grid-based calculations.

It can be seen that for all simulation conditions, the ARO90 basis set provides a better agreement with the numerical reference than the K20 basis set, with the correlation distance being over two times smaller.

In terms of the ionization probability of the hydrogen atom, the ARO90 basis set also provides a good quantitative agreement with the numerical reference, as presented in Fig. 3. The largest discrepancy is observed for $I_0 = 5 \times 10^{14}$ W/cm$^2$ and $n_c = 4$, where the ionization probability at the end of the simulation differs from the grid-based result by about 10%. However, the overall shape of the curve is still reproduced well. The ARO90 basis set performs better than the K20 basis set, especially for the lower two intensities, for which the latter tends to hugely overestimate the ionization probability. This may be attributed to the poor description of the bound excited states by the K20 basis set. The lack of functions with sufficiently large GTO exponents causes the wavefunction to dissipate unphysically and become absorbed by the CAP at a higher rate. For the intensities above the ionization threshold, the results obtained using the two basis sets are comparable, although some details in favor of the ARO20 basis set can be pointed out. The latter provides a better behavior of the ionization probability at the beginning of the simulation, where in the case of the K20 basis set, the norm of the wavefunction begins to decrease too early. Additionally, the constant behavior of the ionization probability at the end of the simulations, observed in both the ARO90 basis set calculations and the grid-based calculations, is not obtained using the K20 basis set.

### C. Toward many-electron cases

To present an example of abilities of the ARO basis sets to describe the electron dynamics in many-electron systems, we performed preliminary real-time time-dependent configuration interaction singles (TD-CIS) calculations for the helium atom subjected to a laser pulse of $I_0 = 2 \times 10^{14}$ W/cm$^2$, $\lambda$ = 800 nm, and four optical cycles and once again compared the performance of an ARO basis set with the performance of a Kauffman basis set. During both fitting procedures, the exponent $\zeta$ was set to 1.7, which is widely accepted as the effective nuclear charge of helium. We use the same quadratic CAP as in the hydrogen atom simulations, with starting position at 1542 bohrs (a value determined using the method described in Sec. IV D). Because in this case, the presence of the electron–electron repulsion necessitates calculations of the two-electron integrals and including the basis functions with all possible values of the azimuthal quantum number, we had to reduce the size of the used basis sets to six lowest angular momenta and to $10 - l$ functions per angular momentum. In the Kauffman scheme, it was achieved by fitting functions only to first ten STO shells (thus, this basis set is dubbed K10$^{He}$). In the ARO scheme, we used a sampling set of the same size as in the calculations of the hydrogen atom but varied the cosine cutoffs to obtain a desired number of GTOs per each angular momentum (thus, this basis set is dubbed ARO90$^{He}$). Both basis sets were supplemented with the 1$s$ contraction optimized to minimize the Hartree–Fock ground state energy of the helium atom.[159] The HHG spectra generated using both basis sets are presented in Fig. 4. As it can be clearly seen, the ARO basis set is able to correctly reproduce the HHG cutoff position, where the K basis set fails completely. The results for even larger systems, including the multicenter ones, are planned to be presented in future works.

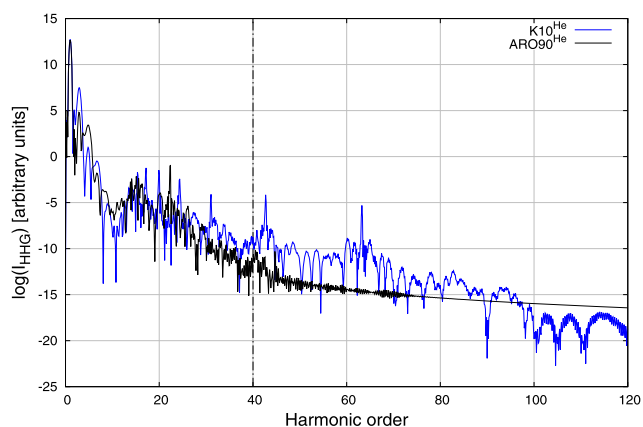

**FIG. 4**. HHG spectra of the helium atom at $I_0 = 2 \times 10^{14}$ W/cm$^2$ and $n_c$ = 4 obtained using the ARO90$^{He}$ basis set and the K10$^{He}$ basis set. The HHG spectra are presented up to thrice the harmonic cutoff value (denoted by the dotted-dashed line).

## VI. CONCLUSION

In this paper, we introduced a novel systematic scheme for the construction of the Gaussian basis sets that are suitable to describe atomic and molecular excited and continuum states. Similar to the approach presented by Kaufmann et al.,[122] our approach has strong theoretical foundations, but it also bypasses most of the limitations related to the use of the K functions. Using the ARO basis sets, we are able to quantitatively or semi-quantitatively reproduce the HHG spectra of the hydrogen atom for intensities below and slightly above the ionization threshold, especially for the shorter exciting pulses. For the intensity well above the ionization threshold and for longer laser pulses, there is some disagreement observed, especially in the harmonic cutoff position. However, in most cases, these discrepancies are notably smaller than using the Kaufmann basis set that has been most frequently used in previous studies covering this subject of research. The ARO basis sets proposed by us also enable a very good description of the ionization probability at practically all intensities.

The results obtained with the ARO basis sets constructed using different reference subsets explicitly show that a large range of Slater orbitals is required to properly approximate the time-dependent wavefunction, even if the laser intensity is relatively low. The omission of STOs with high principal numbers, which is inevitable in the approach proposed by Kauffman et al., leads to a notable worsening of the HHG spectra.

The promising outcomes obtained here for both the hydrogen atom and the helium atom indicate that the ARO basis set can be successfully applied to obtain similar semi-quantitative results for larger systems, which are currently too complex to be captured by more precise yet less affordable grid-based methods. It should be noted that, as demonstrated by the helium atom example, the ARO construction scheme for many-electron atoms remains essentially the same: the sole difference is the STO exponent $\zeta$, which should be replaced by an effective nuclear charge of the atom under consideration.

The basis set approach we explore in this work can be successfully extended to higher levels of theory, such as TD-CISD that allow for examining the effects of electron correlation on attosecond processes. The only complication comes from the rising complexity of the CI matrix (an issue well addressed in the time-independent quantum chemistry) and thus the increasing order of Eq. (16). Moreover, the ARO construction scheme can possibly be adapted to other techniques of solving the TDSE. For instance, when simulating the response in the higher laser frequency regime, the real-time propagation becomes less effective due to the decrease in optimal time step, and alternative approaches that do not require time discretization, such as the $(t, t')$ technique by Moiseyev et al.,[45–47] can be used coupled to the ARO basis sets.

It should be emphasized that in the present paper, we focus on basis sets containing only one type of functions designed to describe the excited and continuum states: either ARO functions or the K functions. We do not rule out the possibility that even better results can be obtained by combining the ARO basis set with some other Gaussian basis sets, analogous to the extensions of the K functions by Luppi et al.[131,135–139] These subjects will also be explored in future works.

This paper also presents a self-consistent method to determine the optimal CAP starting position by utilizing three different approaches: the semiclassical quiver amplitude of the electron in the electromagnetic field, the convergence of the ionization probability, and the optimization of the shape of the HHG spectrum. We also introduce a quantitative measure for comparing the HHG spectra based on the correlation distance that, to our knowledge, is the first to appear in the literature.

## SUPPLEMENTARY MATERIAL

See the supplementary material for the Gaussian basis sets used in the calculations.

## ACKNOWLEDGMENTS

We acknowledge support from the National Science Center, Poland (Symfonia, Grant No. 2016/20/W/ST4/00314). The ICFO group also acknowledges support from ERC AdG NOQIA, Spanish Ministry of Economy and Competitiveness ["Severo Ochoa" program for Centres of Excellence in R&D (Grant No. CEX2019-000910-S)], Plan National FIDEUA (Grant No. PID2019-106901GB-I00/10.13039/501100011033, FPI), Fundació Privada Cellex, Fundació Mir-Puig, Generalitat de Catalunya (AGAUR Grant No. 2017 SGR 1341 and CERCA program Grant No. QuantumCAT_U16-011424, co-funded by the ERDF Operational Program of Catalonia 2014–2020), MINECO-EU QUANTERA MAQS [funded by the State Research Agency (AEI), Grant No. PCI2019-111828-2/10.13039/501100011033], and EU Horizon 2020 FET-OPEN OPTOLogic (Grant No. 899794).

## APPENDIX: CONSTRUCTION OF THE COMPLEX ABSORBING POTENTIAL

Here, we describe a derivation of the complex absorbing potential used in our calculations. The CAP developed by Manolopoulos[149] has a complicated form expressed through the Jacobi elliptic functions, but for practical purposes, it can be approximated as

$$V_{\text{CAP}}^M(r) = E_{\min} \theta(r - r_{\text{CAP}}) y(x), \quad \text{(A1a)}$$

$$y(x) = ax - bx^3 + \frac{4}{(c-x)^2} - \frac{4}{(c+x)^2}, \quad \text{(A1b)}$$

$$x = 2\delta k_{\min}(r - r_{\text{CAP}}). \quad \text{(A1c)}$$

The constants $a$, $b$, and $c$ are defined in Ref. 149. This potential was specifically designed to minimize the transmission and reflection and maximize the absorption of the wavefunction. The absorption efficiency is governed by two parameters, $\delta$ and $k_{\min}$, which are connected to the minimum energy $E_{\min}$ the wavefunction needs to possess in order to be absorbed. Unfortunately, due to a singularity at $x = c$, this potential is applicable only in grid-based approaches, where the singular point is usually placed at the grid boundary.

A frequently used alternative is the so-called monomial CAP

$$V_{\text{CAP}}(r) = \eta \theta(r - r_{\text{CAP}})(r - r_{\text{CAP}})^o \quad \text{(A2)}$$

of order $o$, which is free of singularities, but at the cost of lower flexibility and a less intuitive construction. However, from the perspective of the basis set calculations, another important advantage of the monomial CAP is that the necessary CAP integrals $\langle \chi_{lm;\alpha}^{\text{GTO}} | V_{\text{CAP}} | \chi_{l'm';\alpha'}^{\text{GTO}} \rangle$ can be evaluated analytically. After inserting the general GTO expression (1) into the matrix element and applying the binomial expansion, one obtains a sum of integrals of the form

$$f = N \int_0^\infty \theta(r - r_{\text{CAP}}) r^{2l+o-t+2} \exp(-(\alpha+\alpha')r^2) dr$$
$$= N \int_{r_{\text{CAP}}}^\infty r^{2l+o-t+2} \exp(-(\alpha+\alpha')r^2) dr, \quad \text{(A3a)}$$

$$N = \eta N_\alpha N_{\alpha'} \binom{o}{t} (-r_{\text{CAP}})^t, \quad \text{(A3b)}$$

where $N_\alpha$ and $N_{\alpha'}$ are the GTO normalization constants and $t$ can take on values from 0 to $o$. These integrals have a general solution

$$f = N \frac{r_{\text{CAP}}^{2l+o-t+3}}{2} E_{\frac{1-n}{2}}(r_{\text{CAP}}^2(\alpha+\alpha')), \quad \text{(A4)}$$

where $E_n(z)$ is the generalized exponential integral function.[141]

Because the integrals (A4) have to be evaluated using an extremely high precision, we use the monomial CAP with $o = 2$ to reduce the computational costs. The parameter $\eta$ was determined by performing a series of grid-based calculations and choosing the value that best reproduces the observables: the norm of the wavefunction and the HHG spectra obtained with the Manolopoulos CAP with $\delta = 0.2$ and $k_{\min} = 0.2$. The optimal value was found to be $2.4 \times 10^{-4}$ and proved insensitive to the laser intensity.

It is worth mentioning that although the obtained potential behaves very similar to the Manolopoulos CAP for small $(r - r_{\text{CAP}})$, it rises far more slowly for large $(r - r_{\text{CAP}})$, which may lower the overall absorption efficiency in the grid-based calculations. In order to compensate for it, the width of the absorbing layer (the distance between the CAP starting position and the grid boundary) must be suitably increased. In our calculations, an increase from 32.8 bohrs in the case of the Manolopoulos CAP (a value determined by its parameters) to 100 bohrs in the case of the monomial CAP proved sufficient to reduce any artifacts resulting from partial reflections of the wavefunction (Fig. 5). This problem, however, is absent in the basis set calculations, where the integration of the matrix elements is performed over the whole radial axis.

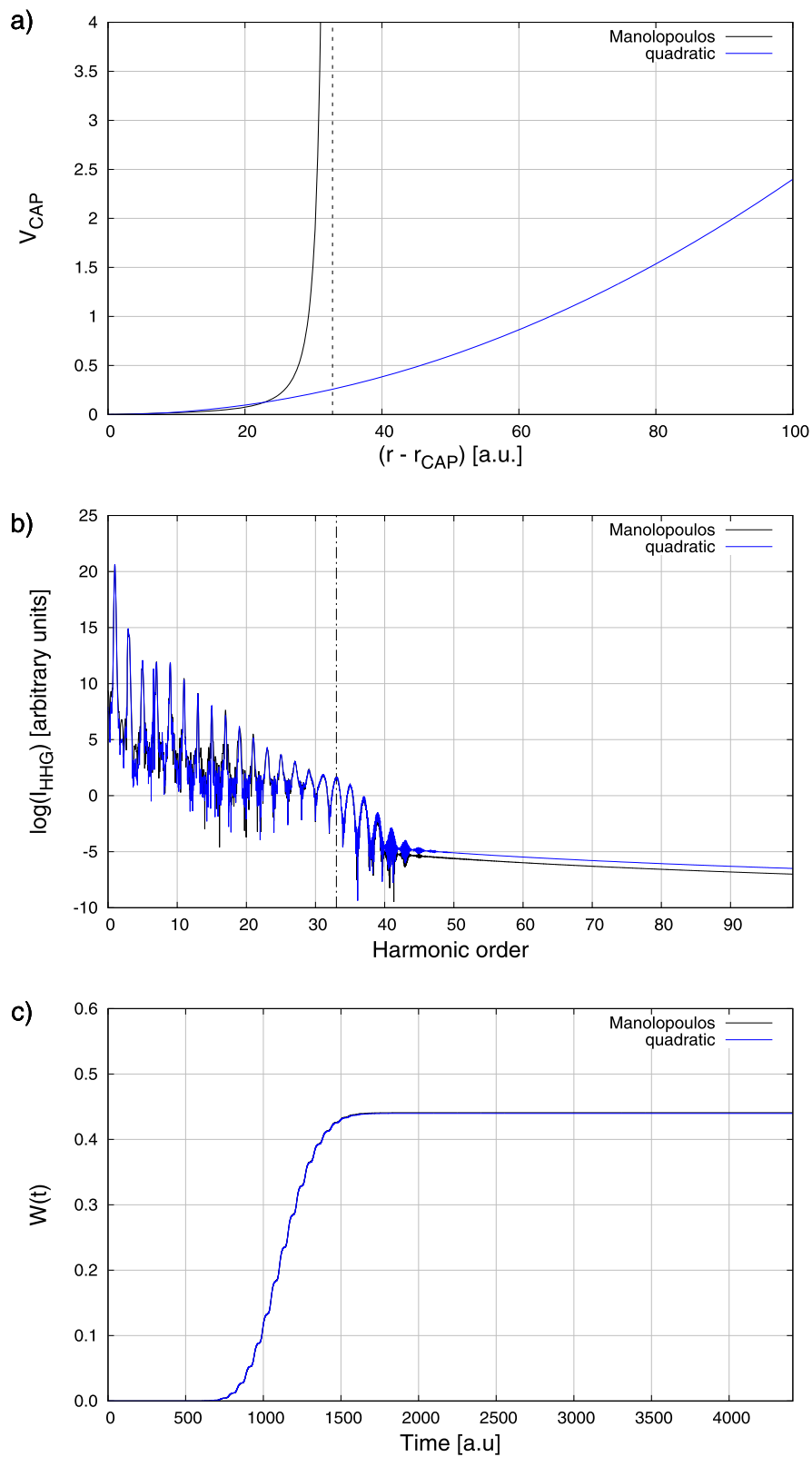

**FIG. 5**. (a) Potential curves of the Manolopoulos CAP and the quadratic CAP employed in our calculations. The singular point of the Manolopoulos CAP is denoted with the dashed line. (b) The HHG spectra of the hydrogen atom obtained with the Manolopulos CAP and the quadratic CAP at $I_0 = 2 \times 10^{14}$ W/cm$^2$ and $n_c = 20$. (c) The time-resolved ionization probability of the hydrogen atom obtained with the Manolopulos CAP and the quadratic CAP at $I_0 = 2 \times 10^{14}$ W/cm$^2$ and $n_c = 20$.

## DATA AVAILABILITY

The data that support the findings of this study are available within the article and its supplementary material.